\documentclass[12pt]{article}

\font\grande=cmr9.5 scaled \magstep4
\font\medio=cmr9.5 scaled \magstep2
\outer\def\beginsection#1\par{\medbreak\bigskip
      \message{#1}\leftline{\bf#1}\nobreak\medskip
\vskip-\parskip
      \noindent}
\usepackage{graphicx} 
\begin{document}
\bibliographystyle {unsrt}

\titlepage

\begin{flushright}
CERN-PH-TH/2013-076
\end{flushright}

\vspace{15mm}
\begin{center}
{\grande Fluid phonons, protoinflationary dynamics}\\
\vspace{5mm}
{\grande and large-scale gravitational fluctuations}\\
\vspace{15mm}
 Massimo Giovannini 
 \footnote{Electronic address: massimo.giovannini@cern.ch} \\
\vspace{0.5cm}
{{\sl Department of Physics, Theory Division, CERN, 1211 Geneva 23, Switzerland }}\\
\vspace{1cm}
{{\sl INFN, Section of Milan-Bicocca, 20126 Milan, Italy}}
\vspace*{1cm}

\end{center}

\vskip 1cm
\centerline{\medio  Abstract}
\vskip 1cm
We explore what can be said on the effective temperature and sound speed of a statistical ensemble of fluid phonons present at the onset of a conventional inflationary phase.  The phonons are the actual normal modes of the gravitating and irrotational fluid that dominates the protoinflationary dynamics. The bounds on the tensor to scalar ratio result in a class of novel constraints involving 
the slow roll parameter, the sound speed of the phonons and the temperature of the plasma prior to the onset of inflation. If the current size of the Hubble radius coincides with the inflationary event horizon redshifted down to the present epoch, the sound speed of the phonons can be assessed from independent measurements of the tensor to scalar ratio and of the tensor spectral index.  
\noindent

\vspace{5mm}

\vfill
\newpage
There are some who argue that the primeval plasma went through a phase 
of quasi-de Sitter expansion whose likelihood is difficult to gauge given our ignorance 
of the preinflationary initial conditions. Since inflation cannot be anyway eternal in its past, 
the inflationary stage of expansion is customarily complemented by 
a preinflationary epoch of decelerated evolution where\footnote{In the present analysis the cosmic or the conformal time parametrizations shall be used indifferently. The derivative with respect to the cosmic time $t$ will be denoted 
by the overdot, while the derivation with respect to the conformal time coordinate $\tau$ will be denoted by a prime. The relation between the parametrizations is given by  $dt = a(\tau)\, d\tau$ where $a(\tau)$ is the scale factor 
of a Friedmann-Robertson-Walker background geometry in the conformal time parametrization. 
Consequently, the Hubble rate  $H = (\dot{a}/a)$ coincides with ${\mathcal H}/a$ where ${\mathcal H} = (a'/a)$. Spatially flat background geometry shall be mainly discussed to make contact with the concordance paradigm but the obtained results can be easily extended to the spatially curved geometries.}
$\dot{a}>0$ and $\ddot{a}<0$. A decelerated phase driven by an irrotational fluid possesses 
its own normal modes that depend on the sound speed of the medium and on the specific coupling to the background geometry.  The quantum theory of the fluctuations of gravitating, irrotational and relativistic fluids \cite{lukash} has been developed even prior to the actual formulation of the conventional inflationary paradigm and in the context of the pioneering analyses of the relativistic theory of large-scale inhomogeneities \cite{lif}. The idea pursued hereunder is, in short, to scrutinize the implications of an ensemble of protoinflationary phonons for the large-scale gravitational perturbations and to compare the obtained results with the 
standard lore provided by the concordance paradigm. 

The preinflationary physics is heuristically modeled by often assuming that the inflaton is the one and only source of the geometry while the initial conditions of the mode functions are taken to differ from the vacuum \cite{standard1} (see also \cite{older1}). We ought first to relax the aforementioned assumption by admitting that initially the total energy-momentum tensor is dominated by an irrotational fluid so that the approximate forms of the scale factors across the protoinflationary transition can be presented in the form of a continuous interpolation\footnote{If the protoinflationary phase is dominated by an irrotational fluid with constant barotropic index $w$, $\alpha = 2/[3(w+1)]$. Note that the interpolation (\ref{proto})--(\ref{inf}) is continuous in $t_{*} = H_{*}^{-1}$, i.e. 
$a_{\mathrm{proto}}(t_{*})= a_{\mathrm{inf}}(t_{*})$ and  $\dot{a}_{\mathrm{proto}}(t_{*})= \dot{a}_{\mathrm{inf}}(t_{*})$.}:
\begin{eqnarray}
&& a_{\mathrm{proto}}(t) = a_{*} (H_{*}\, t)^{\alpha}, \qquad\qquad t < H_{*}^{-1},\qquad 0 \leq \alpha \leq 1,
\label{proto}\\
&& a_{\mathrm{inf}}(t) = a_{*}[ (\alpha/\gamma) H_{*}\, t + (\gamma - \alpha)/\gamma]^{\gamma}, \qquad t > H_{*}^{-1},\qquad \gamma > 1.
\label{inf}
\end{eqnarray}
The time scale $H_{*}^{-1}$ defines the transition between deceleration and acceleration but
three subsequent regimes can be physically distinguished: first the preinflationary epoch (where the fluid dominate); second the protoinflationary regime (where $\ddot{a}$ changes sign); third the fully developed inflationary stage. Defining $\epsilon = - \dot{H}/H^2$ as the rate of variation of the Hubble parameter, we shall have that $\epsilon\simeq 3(w+1)/2>1$ during the first regime. Conversely $\epsilon \simeq {\mathcal O}(1)$ during the protoinflationary phase and $\epsilon \ll 1$ as soon as the slow-roll evolution starts.  Explicit examples of this dynamics 
can be found in terms of either analytic or numerical solutions of the Einstein equations with multiple sources. 
Sticking to a simplistic model, the interpolating scale factor $a(t) = a_{*} (\sinh{\beta H_{*}})^{1/\beta}$ (with $\beta = 3(w+1)/2$) is a solution of the Friedmann equations supplemented by a perfect barotropic fluid and and by a minimally coupled scalar field with inflating potential; in this case  $\epsilon(t) = \beta/\cosh^2{(\beta H_{*} t)}$. 

During the protoinflationary phase the fluctuations of an irrotational and perfect fluid are  
described in terms of an action whose explicit form has been derived, for the first time, by Lukash \cite{lukash}
\begin{equation}
S_{\mathrm{phonons}} = \frac{1}{2}\int d^{4} x\, \biggl[ (\partial_{\tau} q_{\mathrm{ph}})^2
- c_{\mathrm{s}}^2(\tau) \gamma^{ij}\partial_{i}  q_{\mathrm{ph}}
\partial_{j} q_{\mathrm{ph}}  + \frac{z_{\mathrm{ph}}''}{z_{\mathrm{ph}}} 
q_{\mathrm{ph}}^2\biggr], \qquad z_{\mathrm{ph}} = \frac{a^2 \sqrt{p_{\mathrm{t}} + \rho_{\mathrm{t}}}}{ c_{\mathrm{s}} {\mathcal H}},
\label{phon1}
\end{equation}
where $\gamma^{ij}\equiv \delta^{ij}$ in the spatially flat case and 
$c_{\mathrm{s}}^2(\tau) = p_{\mathrm{t}}'/\rho_{\mathrm{t}}'$ denotes the total sound speed of the system measured in natural units.  The variable $q_{\mathrm{ph}}$ of Ref. \cite{lukash}  can be expressed in terms of ${\mathcal R}$, i.e. the  
curvature perturbations on comoving orthogonal hypersurfaces satisfying, because of the action (\ref{phon1}), the following equation: 
\begin{equation}
{\mathcal R}'' + 2 \frac{z_{\mathrm{ph}}'}{z_{\mathrm{ph}}} {\mathcal R}' - c_{\mathrm{s}}^2 \,\nabla^2 {\mathcal R} =0, \qquad {\mathcal R} = - q_{\mathrm{ph}}/z_{\mathrm{ph}}.
\label{phon1a}
\end{equation}

Besides the fluid phonons the other modes of vibrations of the protoinflationary fluid are associated with 
spin-two inhomogeneities \cite{lif} whose action follows from the second-order fluctuation of the Einstein-Hilbert action in the tensor amplitude 
$h_{ij}$ (with $\partial_{i} h^{ij}= \, h_{i}^{i} =0$),
\begin{equation}
S_{\mathrm{gravitons}} = \frac{1}{8\ell_{\mathrm{P}}^2} \int d^{4} x \, a^2(\tau)  \,\biggl[ 
\partial_{\tau} h_{ij} \partial_{\tau} h^{ij} - \gamma^{k \ell}\partial_{k} h_{ij} \partial_{\ell} h^{ij} \biggr],
\qquad \ell_{\mathrm{P}} = \sqrt{8\pi G} =\frac{8\pi}{M_{\mathrm{P}}}= 
\frac{1}{\overline{M}_{\mathrm{P}}},
\label{EQ3}
\end{equation}
where $\ell_{\mathrm{P}}$ denotes the Planck length, $M_{\mathrm{P}}$ is the Planck mass 
and $\overline{M}_{\mathrm{P}}$ is the reduced Planck mass.
The variables $h_{ij}$, ${\mathcal R}$ and $q_{\mathrm{ph}}$ are gauge-invariant and can be evaluated in the coordinate system which is more suitable for the problem at hand. This statement holds, of course, for the scalar 
modes since the tensors are automatically gauge-invariant. Using, for instance, the off diagonal gauge \cite{OD} (where  $\delta_{\mathrm{s}} g_{00} =2 a^2 \phi$ and $\delta_{\mathrm{s}} g_{0i} = - a^2 \beta$) the fluctuations of the full metric are:
\begin{eqnarray}
&& \delta_{\mathrm{s}} g_{00} (\vec{x},\tau)= \frac{a^2(\tau) \,{\mathcal H}^2(\tau)}{{\mathcal H}'(\tau) - {\mathcal H}^2(\tau)} {\mathcal R}(\vec{x},\tau),
\nonumber\\
&& \delta_{\mathrm{s}} g_{0i}(\vec{x},\tau) = 
- \frac{a^2(\tau)}{{\mathcal H}(\tau)}{\mathcal R}(\vec{x},\tau) + \int^{\tau} a^2(\tau_{1}) \, {\mathcal R}(\vec{x},\tau_{1})\, d\tau_{1} 
+ \int^{\tau} \frac{a^2(\tau_{1})}{ {\mathcal H}(\tau_{1})} {\mathcal R}'(\vec{x}, \tau_{1}) \, d\tau_{1},
\label{phon2}\\
&& \delta_{\mathrm{t}} g_{ij}(\vec{x},\tau) = - a^2(\tau) \,\, h_{ij}(\vec{x},\tau).
\label{phon3}
\end{eqnarray}
In Eqs. (\ref{phon2})--(\ref{phon3}) the tensor and scalar perturbations are denoted, respectively, by  $\delta_{\mathrm{t}}$ and
 $\delta_{\mathrm{s}}$; $\delta_{\mathrm{s}} g_{ij}(\vec{x},\tau)$ vanishes in the off diagonal gauge \cite{OD}. 
Equations (\ref{phon2}) follow from the perturbed Einstein equations implying, in the absence of anisotropic stresses, 
$\beta' + 2 {\mathcal H}\beta + \phi=0$; the Bardeen potential \cite{bard1}, in the off-diagonal gauge, can be 
expressed as $\Psi(\vec{x},\tau) = - {\mathcal H}(\tau) \beta(\vec{x},\tau)$.

After promoting the normal mode  $q_{\mathrm{ph}}(\vec{x},\tau)$ and $h_{ij}(\vec{x},\tau)$  to the status of field operators obeying (equal time) canonical commutation relations with their conjugate momenta, the mode expansions of the corresponding operators become:
\begin{eqnarray}
&& \hat{q}_{\mathrm{ph}}(\vec{x},\tau) =\int\,\,  \frac{ d^{3}k }{(2 \pi)^{3/2}} \,\,\biggl[ f_{k}(\tau) \, \hat{a}_{\vec{k}} \, e^{- i \, \vec{k}\cdot \vec{x}} 
+ f_{k}^{*}(\tau) \, \hat{a}_{\vec{k}}^{\dagger} \, e^{ i \, \vec{k}\cdot \vec{x}} \biggr],
\label{Qph}\\
&& \hat{h}_{ij}(\vec{x},\tau)= \frac{\sqrt{2} \ell_{\mathrm{P}}}{a(\tau)}\sum_{\lambda}
  \int \, \frac{d^{3}k}{(2\pi)^{3/2}}\,\, \epsilon_{ij}^{(\lambda)}(\hat{k})\, 
 \biggl[ \hat{b}_{\vec{k},\lambda} \,g_{k,\lambda}(\tau) e^{- i \vec{k} \cdot \vec{x}} +  \hat{b}_{\vec{k},\lambda}^{\dagger}\, g_{k,\lambda}^{*}(\tau) \, e^{ i \vec{k} \cdot \vec{x}}\biggr],
 \label{Qgrav}
 \end{eqnarray}
where $[\hat{b}_{\vec{k},\lambda}, \hat{b}^{\dagger}_{\vec{p},\lambda'}] = \delta^{(3)}(\vec{k} - \vec{p})
\delta_{\lambda\lambda'}$ and $[\hat{a}_{\vec{k}}, \, \hat{a}_{\vec{p}}^{\dagger}] = \delta^{(3)}(\vec{k} - \vec{p})$;
$ \epsilon_{ij}^{(\lambda)}(\hat{k})$ accounts for the two propagating polarizations of the tensor modes and satisfies 
$\epsilon_{ij}^{(\lambda)} \epsilon_{ij}^{(\lambda')} = 2 \delta_{\lambda\lambda'}$.
The mode functions $f_{k}$  and $g_{k,\lambda}$ obey two similar equations leading, however, to different dispersion relations:
\begin{equation}
 f_{k}'' + \biggl[ k^2 \, c_{\mathrm{s}}^2 - \frac{z_{\mathrm{ph}}''}{z_{\mathrm{ph}}}\biggr] f_{k} =0, \qquad 
g_{k,\lambda}'' + \biggl[ k^2 - \frac{a''}{a}\biggr] g_{k,\lambda} =0.
\label{Q2}
\end{equation}
During the protoinflationary regime $\epsilon$ decreases from values $\epsilon(t) > 1$ to values $\epsilon(t) \simeq 
{\mathcal O}(1)$. As soon as $\epsilon(t) \ll 1$, the evolution 
of curvature perturbations is governed virtually by Eq. (\ref{infl1}) 
except for the dispersion relations (implying $c_{\mathrm{s}} \to 1$) and except for a different form of the pump field 
i.e.\footnote{Equation (\ref{infl1}) can be derived within the same logic employed in the first paper of
Ref. \cite{lukash} but in the case of scalar field matter \cite{KS}.}
\begin{equation}
{\mathcal R}'' + 2 \frac{z_{\varphi}'}{z_{\varphi}} {\mathcal R}' - \nabla^2 {\mathcal R} =0, 
\qquad z_{\varphi} = \frac{a \varphi'}{{\mathcal H}},\qquad q_{\varphi} = - z_{\varphi} \, {\mathcal R},
\label{infl1}
\end{equation}
where $\varphi$ denotes the inflaton field and $q_{\varphi}$ is just the analog of $q_{\mathrm{ph}}$ of Eq. (\ref{phon1}). 
As discussed after Eqs. (\ref{proto})--(\ref{inf}) a smooth protoinflationary transition implies the smoothness of the scale factors,  of the curvature pump  fields 
(i.e.$z_{\mathrm{ph}}$ and $z_{\varphi}$) and, ultimately,
of the extrinsic curvature. The inhomogeneities can be matched by requiring the continuity of the induced three metric and of the extrinsic curvature on the space-like hypersurface of the transition \cite{trans}. Such a requirement demands, at the level of  Eqs. (\ref{phon2})--(\ref{phon3}), the continuity of ${\mathcal R}(\vec{x},\tau)$, $\beta(\vec{x},\tau)$ and of $h_{ij}(\vec{x},\tau)$. The results obtained in this way
can be directly checked by deriving the evolution equations during a phase where the fluid phonons and the inflaton quanta are simultaneously present but the fluid component is already subleading (see e.g. last paper of Ref. \cite{standard1}).

Since the phonons and the gravitons evolve, initially, in a fluid background (which may even be dominated by radiation) the initial state for each species must be 
reasonably described by a mixed density matrix with Bose-Einstein distribution i.e. 
\begin{equation}
\hat{\rho} = \sum_{\{n\}} P_{\{n \}} |\{n \}\rangle \langle \{n \}|,
\qquad P_{\{n\}} = \prod_{\vec{k}} \frac{\overline{n}_{k}^{n_{k}}}{( 1 + \overline{n}_{k})^{n_{k} + 1 }},
\label{matrix1}
\end{equation}
where $\overline{n}_{k}$ is the average multiplicity of each Fourier mode and $ |\{n \}\rangle = |n_{\vec{k}_{1}} \rangle \otimes  |n_{\vec{k}_{2}} \rangle \otimes  |n_{\vec{k}_{3}} \rangle...$ with the ellipses standing for all the occupied modes of the field. The average multiplicities of gravitons and phonons are given by: 
\begin{equation}
\overline{n}^{\mathrm{ph}}_{k} = \frac{1}{e^{\omega^{\mathrm{ph}}/k_{\mathrm{T}}}-1}, \qquad \overline{n}^{\mathrm{gr}}_{k} = \frac{1}{e^{\omega^{\mathrm{gr}}/k_{\mathrm{T}}} -1},
\label{occup}
\end{equation}
having denoted with $k_{\mathrm{T}} = T$ the effective temperature of the phonons and of the gravitons\footnote{We shall posit that phonons and gravitons have the same temperature but this assumption can be dropped if only approximate (kinetic) equilibrium holds between 
the different species.}. Note that $\omega^{\mathrm{ph}}= k c_{\mathrm{s}}$ while $\omega^{\mathrm{gr}}= k$. Multiplying the occupation numbers of Eq. (\ref{occup})
by the corresponding quanta of energy and integrating the result over the appropriate phase space, the energy density of phonons and gravitons 
becomes $g_{\mathrm{eff}}\pi^2 T_{*}^4/30$ where $g_{\mathrm{eff}}$ denotes the effective number of relativistic degrees of freedom. If only gravitons and phonons are considered $g_{\mathrm{eff}} =3$ but
$g_{\mathrm{eff}}$ maybe larger than $3$ because of other species in local thermal equilibrium.
If  $T_{\mathrm{max}} = [ 45/(4\pi ^3 g_{\mathrm{eff}})]^{1/4} 
\sqrt{H_{*} M_{\mathrm{P}}}$ the energy density associated with the initial state is ${\mathcal O}( 3 H_{*}^2 M_{\mathrm{P}}^2)$. For the forthcoming applications it will be practical to define $T_{*} = Q\, T_{\mathrm{max}}$ where $Q\leq 1$ measures the protoinflationary 
temperature in units of $T_{\mathrm{max}}$.

The initial mixed quantum state of the system is the direct product of the quantum state of the phonons and of the gravitons, i.e. $|\Phi \rangle = |\psi_{\mathrm{phonons}}\rangle\, \otimes\, |\phi_{\mathrm{gravitons}}\rangle$ and the corresponding power spectra are obtained from the following two expectation values: 
\begin{eqnarray}
&& \langle \Phi | \hat{{\mathcal R}}(\vec{x}, \tau) \hat{{\mathcal R}}(\vec{x} + \vec{r}, \tau) | \Phi \rangle = 
\int \frac{d k}{k} {\mathcal P}_{{\mathcal R}}(k,\tau) \frac{\sin{k r}}{k r},
 \label{SP1}\\
 && \langle \Phi | \hat{h}_{ij}(\vec{x}, \tau) \hat{h}^{ij}(\vec{x} + \vec{r}, \tau) | \Phi \rangle = 
 \int \frac{d k}{k} {\mathcal P}_{{\mathrm{T}}}(k,\tau) \frac{\sin{k r}}{k r},
\label{SP2}
\end{eqnarray}
where\footnote{We used, in Eq. (\ref{SP2}) that the mode function of each tensor polarization is the same since the two polarizations 
do not interact between them. } the two power spectra are:
\begin{equation} 
{\mathcal P}_{{\mathcal R}}(k,\tau) = \frac{k^3}{2\pi^2} |f_{k}(\tau)|^2 ( 2 \overline{n}^{\mathrm{ph}}_{k} +1),\qquad 
{\mathcal P}_{{\mathrm{T}}}(k,\tau) = \frac{4 \ell_{\mathrm{P}}^2}{a^2 \, \pi^2} \, k^3\, |g_{k}(\tau)|^2 ( 2 \overline{n}^{\mathrm{gr}}_{k} +1).
\label{SP3}
\end{equation}
These spectra must be evaluated in the limits $k/(a_{*} H_{*}) \gg 1$ (corresponding to wavelengths smaller then the Hubble radius at the protoinflationary transition) and $k/(a H) \ll 1$  (since the phenomenologically relevant wavelengths are still larger than the Hubble radius at the epoch of the matter radiation equality). In these concurrent limits Eq. (\ref{SP3}) implies
\begin{eqnarray}
{\mathcal P}_{{\mathcal R}}(k,\tau) &=& \frac{8}{3 M_{\mathrm{P}}^4} \biggl(\frac{V}{\epsilon}\biggr) \biggl(\frac{k}{a H}\biggr)^{n_{\mathrm{s}} -1} \,{\mathcal F}_{{\mathcal R}}(k,\, \tau_{*},\, c_{\mathrm{s}},\,
\epsilon,\, \eta) \, \coth{\biggl(\frac{k c_{\mathrm{s}}}{2 k_{\mathrm{T}}}\biggr)},
\label{SP4}\\
{\mathcal P}_{\mathrm{T}}(k,\tau) &=& \frac{128}{3 } \biggl(\frac{V}{M_{\mathrm{P}}^4}\biggr) \biggl(\frac{k}{a H}\biggr)^{n_{\mathrm{T}}} \,{\mathcal F}_{\mathrm{T}}(k,\, \tau_{*},\, c_{\mathrm{s}},\,
\epsilon) \, \coth{\biggl(\frac{k}{2 k_{\mathrm{T}}}\biggr)},
\label{SP5}
\end{eqnarray}
where $\eta= \ddot{\varphi}/(H \dot{\varphi}) = (\epsilon - \overline{\eta})$ and $V$ is the inflaton potential; the two spectral indices are:
\begin{equation}
n_{\mathrm{s}} = 1 - 6 \epsilon + 2 \overline{\eta}, \qquad n_{\mathrm{T}} = - 2 \epsilon.
\label{SP6}
\end{equation}
For $k/(a_{*} H_{*}) \gg 1$ the two functions ${\mathcal F}_{{\mathcal R}}$ and ${\mathcal F}_{\mathrm{T}}$ go to $1$ up to corrections ${\mathcal O}(k^{-2}\tau_{*}^{-2})$ which will be irrelevant for the subsequent discussion.  From Eqs. (\ref{SP4}) and (\ref{SP5})  the tensor 
to scalar ratio is given by 
\begin{equation}
r_{\mathrm{T}}(k,\epsilon,c_{\mathrm{s}}, k_{\mathrm{T}}) = 16 \epsilon \biggl(\frac{k}{k_{\mathrm{p}}}\biggr)^{1 - n_{\mathrm{s}} - n_{\mathrm{T}}} 
 \frac{\coth{[k /(2 k_{\mathrm{T}})]}}{\coth{[k\, c_{\mathrm{s}}/(2 k_{\mathrm{T}})]}},
 \label{RATIO}
 \end{equation}
where the power spectra have been referred to the standard pivot scale $k_{\mathrm{p}} =0.002\,\, \mathrm{Mpc}^{-1}$ corresponding to an effective multipole $\ell_{\mathrm{eff}} \simeq 30$. 

In the concordance paradigm  radiation takes places almost suddenly after inflation and
long phases (different from radiation) are excluded down to the curvature scale of matter-radiation equality. 
Under this approximation the ratio $(k/k_{\mathrm{T}})$ is given by
\begin{equation}
\frac{k}{k_{\mathrm{T}}} = \frac{\pi}{Q} g_{\mathrm{eff}}^{1/4}\, \biggl(\frac{4 {\mathcal A}_{{\mathcal R}}\, \epsilon}{45} \biggr)^{1/4} \biggl(\frac{k}{H_{0}}\biggr) \, e^{ - (N_{\mathrm{max}} - N)}, 
\label{num1}
\end{equation}
where $N_{\mathrm{max}}$ denotes the maximal number of efolds presently accessible through large-scale 
observations and ${\mathcal A}_{R}$ is the amplitude 
of the power spectrum of curvature perturbations determined at the pivot scale $k_{\mathrm{p}}$. In practice $N_{\mathrm{max}}$ is determined by fitting inside 
the Hubble radius the approximate size of the inflationary event horizon redshifted down to the present epoch:
\begin{equation}
N_{\mathrm{max}} = 62.2 + 0.5 \ln{( 10^{5} \, \xi)} - \ln{(h_{0}/0.7)} + 0.25 \ln{[10^{5}\, h_{0}^2 \Omega_{\mathrm{R}0}/4.15]},
\label{num2}
\end{equation}
where $\xi = \pi \epsilon {\mathcal A}_{{\mathcal R}}$ and $\Omega_{\mathrm{R}0}$ is the present critical 
fraction of radiation energy density. If the  postinflationary 
phase contains long epochs expanding with a rate which is slower than radiation  $N_{\mathrm{max}}$ can even be 
${\mathcal O}(15)$ efolds larger.

The WMAP9 yr data (first and second papers of Ref. \cite{wmap}) alone imply $r_{\mathrm{T}}(k_{\mathrm{p}}) <0.38$. 
When the WMAP9 data are  combined with other data sets\footnote{In particular the data on the Hubble rate, the ones on the baryon acoustic oscillations supplemented by the data of the Atacama Cosmology Telescope, by the data of the south pole telescope and by the three year sample of the supernova legacy survey (see the papers reported in Ref. \cite{other})} we have that 
the cosmological parameters are determined to be 
\begin{equation}
( \Omega_{\mathrm{b}0}, \, \Omega_{\mathrm{c}0}, \Omega_{\mathrm{de}0},\, h_{0},\,n_{\mathrm{s}},\, \epsilon_{\mathrm{re}}) \equiv 
(0.04596,\, 0.2360,\, 0.7180,\,0.6969,\, 0.9647,\,0.081),
\label{par}
\end{equation}
while ${\mathcal A}_{\mathcal R} = 2.431\times 10^{-9}$. In the case of the parameters of Eq. (\ref{par}) the bounds on $r_{\mathrm{T}}$ and on the 
tensor spectral index imply, respectively,  $r_{\mathrm{T}}(k_{\mathrm{p}}) < 0.13$ (95 \% CL) and $n_{\mathrm{T}} > -0.016$. The Planck satellite gives $r_{\mathrm{T}}(k_{\mathrm{p}}) <0.11$ (see the last paper of Ref. \cite{other}) which is quantitatively comparable with the WMAP9 limit.

If $N \gg N_{\mathrm{max}}$ Eqs. (\ref{RATIO}) and (\ref{num1})--(\ref{num2}) imply that 
 $r_{\mathrm{T}}(k_{\mathrm{p}}) \to 16 \epsilon$. Similarly, for $c_{\mathrm{s}} \to 1$
(i.e. when the sound speed coincides with the speed of light) $r_{\mathrm{T}}(k_{\mathrm{p}}) = 16 \epsilon$ in spite of the total number of efolds. This is the so-called consistency regime stipulating that, at the pivot scale, $r_{\mathrm{T}} \simeq 16\epsilon \simeq - 8 n_{\mathrm{T}}$. Whenever the speed of sound does not equal the speed of light the consistency relation is violated.
\begin{figure}[!ht]
\centering
\includegraphics[height=5cm]{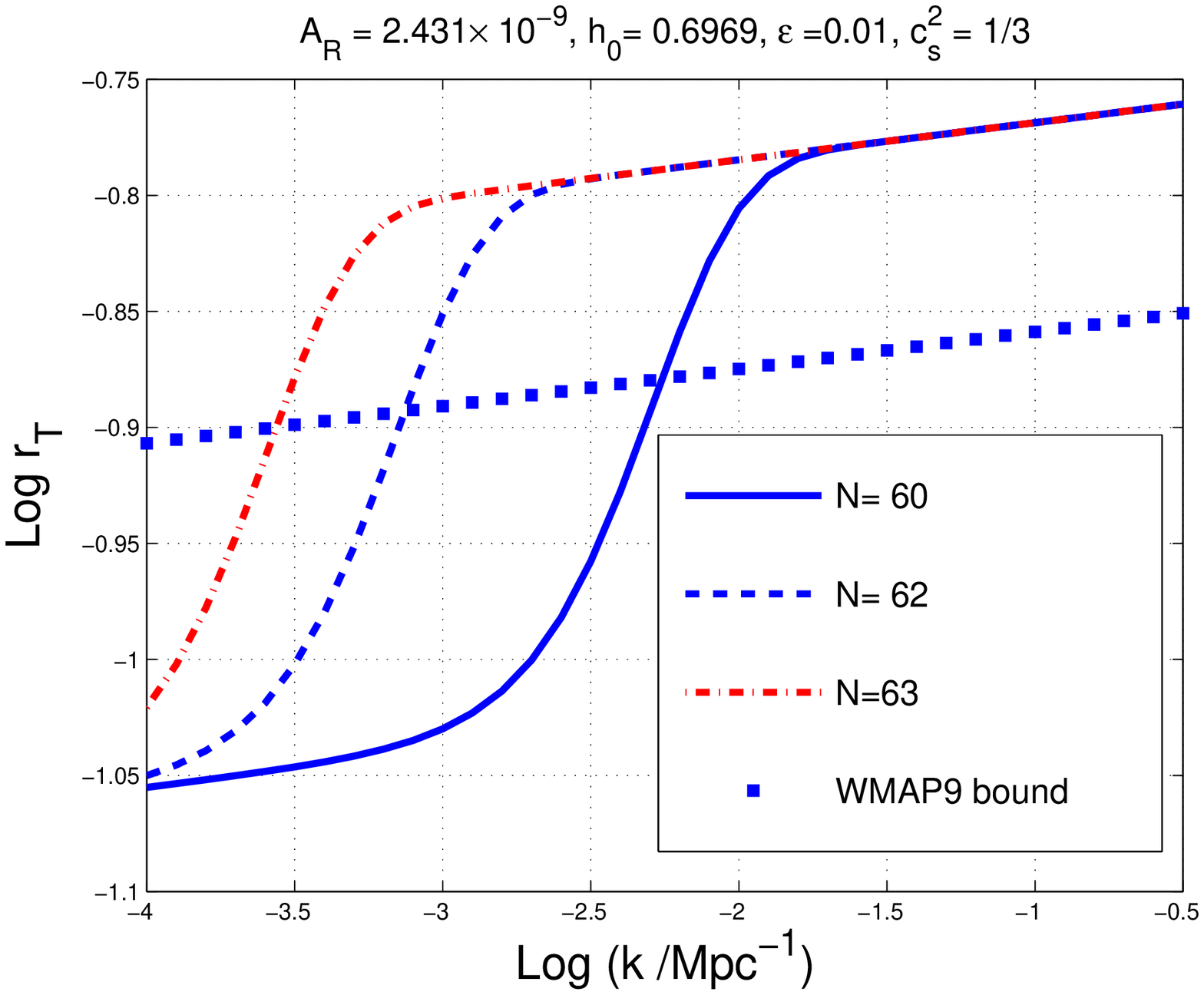}
\includegraphics[height=5cm]{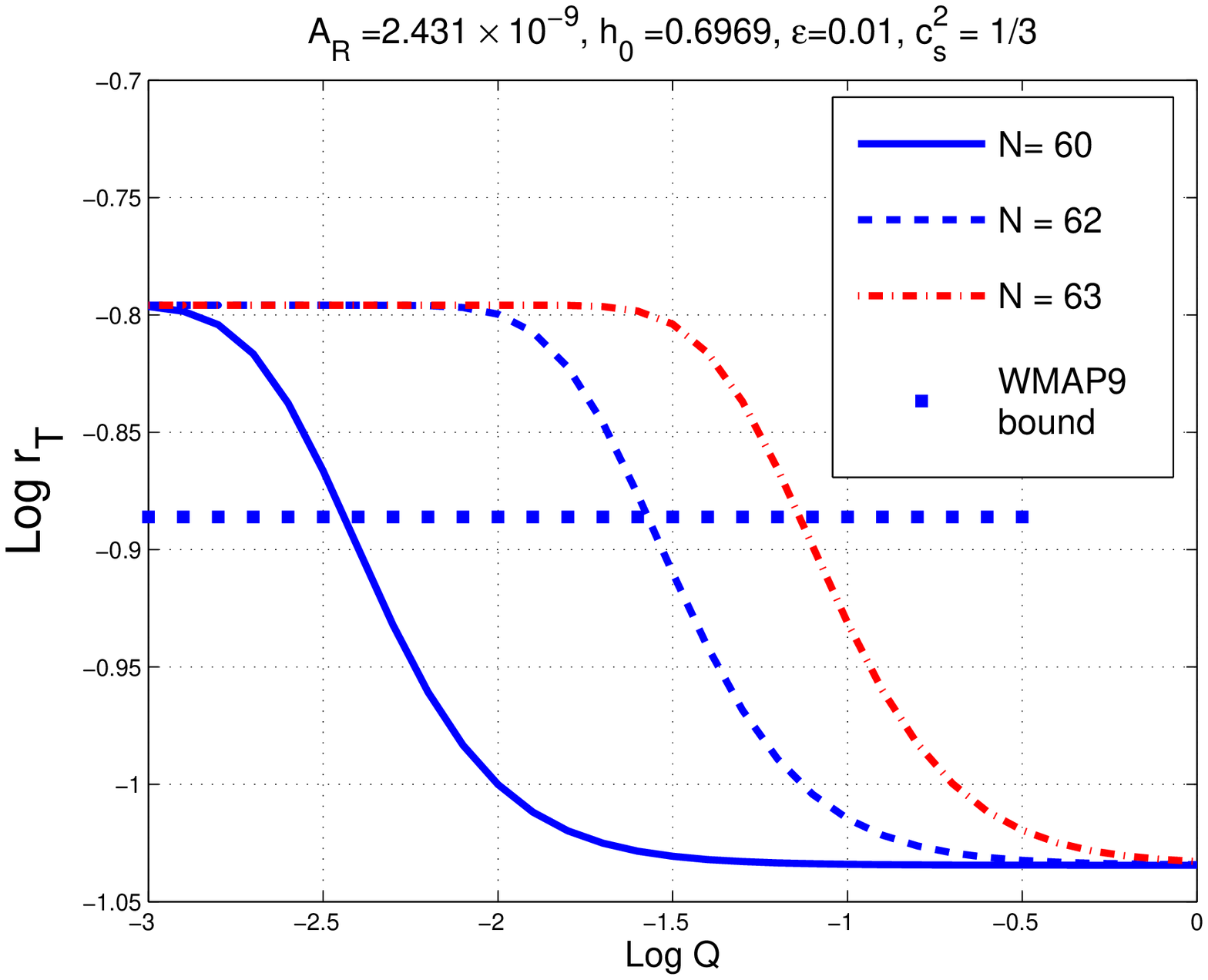}
\caption[a]{The tensor to scalar ratio for the fiducial set of parameters of Eq. (\ref{par}). In the left plot $Q=0.01$ and $k$ is allowed to vary; in the right 
plot $k=k_{\mathrm{p}}$. On the vertical and the horizontal axes the common logarithms of the corresponding 
quantities are illustrated.}
\label{Figure}      
\end{figure}
According to Eqs. (\ref{num1})--(\ref{par}), if  $N\simeq {\mathcal O}(N_{\mathrm{max}})$ then $k/k_{\mathrm{T}} \simeq {\mathcal O}(1)$ as long as $60\leq N\leq64$,  $0.01\leq Q\leq 1$ and 
 $0.26 g_{\mathrm{eff}}^{1/4} \leq (k_{\mathrm{p}}/k_{\mathrm{T}})\leq 0.48 g_{\mathrm{eff}}^{1/4}$. Thus
$r_{\mathrm{T}}(k_{\mathrm{p}},\epsilon,c_{\mathrm{s}}, k_{\mathrm{T}})$  can be expanded for $ k_{\mathrm{p}}/(2k_{\mathrm{T}}) < 1$ so that, approximately, $r_{\mathrm{T}} \simeq 16 \,\epsilon \, c_{\mathrm{s}}(T_{\mathrm{gr}}/T_{\mathrm{ph}})$ where the case of different (kinetic) temperatures has been included for completeness (in the thermal equilibrium case $T_{\mathrm{ph}}\simeq T_{\mathrm{gr}}$).  This means that the consistency relation is violated and the sound speed 
of the primordial phonons can be potentially measured from independent estimates of the tensor to scalar ratio and of the tensor spectral index since it is 
still true that $n_{\mathrm{T}} = - 2 \epsilon$.
In Fig. \ref{Figure} the value of $r_{\mathrm{T}}$ is illustrated without approximations and for different values of the inflationary efolds. In the left plot  $Q$ is held fixed while in the right plot $k$ coincides with $k_{\mathrm{p}}$. In both plots the sound speed has been chosen to equal the radiation value, i.e. $c_{\mathrm{s}}^2 =1/3$. With the squares we illustrate the WMAP9 limit on $r_{\mathrm{T}}$. 
The allowed region is always below the squares. The WMAP9 limit is purely illustrative since it is derived, as usual, by assuming the consistency relation.

All in all the results of this investigation imply that a protoinflationary phase dominated by fluid phonons 
invalidates the consistency paradigm by impacting on the tensor to scalar ratio whose 
spectral dependence is determined by the sound speed of the phonons and by the potentially different temperatures of the various species in kinetic equilibrium. 
Reversing the argument, the present findings show more modestly that the consistency relations, even if 
heuristically assumed in nearly all experimental analyses, are not always tenable and bear the mark of an essential 
completion of the early inflationary dynamics, i.e. the protoinflationary transition. It is therefore tempting to speculate that independent measurements of $r_{\mathrm{T}}$ and $n_{\mathrm{T}}$ from B-mode polarization may offer a potentially novel diagnostic of the role played by primordial phonons in setting the initial conditions of large-scale gravitational perturbations.

\end{document}